\documentclass[11pt,aps,
superscriptaddress,prc,eprint,nofootinbib]{revtex4-2}
\pdfoutput=1
\usepackage{graphicx}
\usepackage{dcolumn}
\usepackage{bm}

\usepackage{verbatim}
\usepackage{empheq}

\usepackage{amsmath,amsfonts,amsthm,amssymb}
\usepackage{graphics}
\usepackage{cancel}
\usepackage{multirow}
\usepackage{longtable}
\usepackage{color}
\usepackage{xcolor}
\usepackage[normalem]{ulem}

\usepackage{dsfont}

\newcounter{rownum}
\usepackage{amssymb}

\usepackage{mathtools}
\usepackage{subfigure}
\usepackage{bbold}
\usepackage{yfonts}
\usepackage[colorlinks=true,linktocpage=true,linkcolor=blue,citecolor=blue]{hyperref}
\usepackage{placeins}
\usepackage{bm} 
\usepackage{nicefrac}
\usepackage{slashed}
\usepackage{marginnote}

\newcommand{\bea}{\begin{eqnarray}}
\newcommand{\eea}{\end{eqnarray}}
\newcommand{\bel}[1]{\begin{eqnarray}\label{#1}}
\newcommand{\eel}{\end{eqnarray}}
\newcommand{\beq}{\begin{equation}}
\newcommand{\eeq}{\end{equation}}

\def\LB{\left(}
\def\RB{\right)}
\def\LSB{\left[}
\def\RSB{\right]}

\newcommand{\f}[2]{\frac{#1}{#2}}
\newcommand{\onehalf}{{\nicefrac{1}{2}}} 
 
\newcommand{\threefourths}{{\nicefrac{3}{4}}} 

\newcommand{\dd}{\mathrm{d}}
\def\spin{\,\textgoth{s:}}
\def\spinl{|{\boldsymbol s}_*|}

\newcommand{\EQ}[1]{Eq.~(\ref{#1})}
\newcommand{\EQn}[1]{(\ref{#1})}

\newcommand{\CITn}[1]{\citep{#1}} 

\newcommand{\av}{{\boldsymbol a}} 
\newcommand{\bv}{{\boldsymbol b}}

\newcommand{\nv}{{\boldsymbol n}}

\newcommand{\pv}{{\boldsymbol p}}

\newcommand{\sv}{{\boldsymbol s}}

\newcommand\alphav{{\boldsymbol \alpha}}

\definecolor{amethyst}{rgb}{0.6, 0.4, 0.8}

\begin{document}
	
\title{Nonlinear causality and stability of perfect spin hydrodynamics and its nonperturbative character}

\author{Samapan Bhadury}
\email{bhadury.samapan@gmail.com}
\affiliation{Institute of Theoretical Physics, Jagiellonian University, ul. St. \L ojasiewicza 11, 30-348 Kraków, Poland}

\author{Zbigniew Drogosz}
\email{zbigniew.drogosz@alumni.uj.edu.pl}
\affiliation{Institute of Theoretical Physics, Jagiellonian University, ul. St. \L ojasiewicza 11, 30-348 Kraków, Poland}

\author{Wojciech Florkowski}
\email{wojciech.florkowski@uj.edu.pl}
\affiliation{Institute of Theoretical Physics, Jagiellonian University, ul. St. \L ojasiewicza 11, 30-348 Kraków, Poland}

\author{Sudip Kumar Kar}
\email{sudip.kar@doctoral.uj.edu.pl}
\affiliation{Institute of Theoretical Physics, Jagiellonian University, ul. St. \L ojasiewicza 11, 30-348 Kraków, Poland}

\author{Valeriya Mykhaylova}
\email{valeriya.mykhaylova@uj.edu.pl}
\affiliation{Institute of Theoretical Physics, Jagiellonian University, ul. St. \L ojasiewicza 11, 30-348 Kraków, Poland}

\date{\today}

\begin{abstract}
Four formulations of perfect spin hydrodynamics for spin-$\onehalf$ particles, distinguished by their treatment of spin (classical vs.~quantum) and by the underlying particle statistics (Boltzmann vs.~Fermi--Dirac), are analyzed and shown to satisfy the requirements of a divergence-type theory. Moreover, for all the formulations, we define the generating functions associated with the relevant thermodynamic currents and demonstrate that the constructed hydrodynamic theory is nonlinearly causal and stable. The latter is achieved by employing the exact expressions for the distribution functions, indicating a nonperturbative character of our approach. 
\end{abstract}

\maketitle
\bigskip
\section{Introduction}

The recent measurements showing that the particles produced in relativistic heavy-ion collisions exhibit nontrivial spin polarization~\cite{STAR:2017ckg, STAR:2018gyt, STAR:2019erd, ALICE:2019aid,STAR:2025jhp,STAR:2022fan,Li:2025awj, STAR:2025jwc, Shen:2024rdq} have sparked significant interest in the development of spin hydrodynamics~\mbox{\cite{Becattini:2021iol, Palermo:2024tza,Sheng:2025cjk,Becattini:2025oyi,Florkowski:2017ruc, Florkowski:2017dyn, Florkowski:2018ahw, Florkowski:2018fap, Bhadury:2020puc, Bhadury:2022ulr, Weickgenannt:2019dks, Weickgenannt:2021cuo, Weickgenannt:2020aaf, Weickgenannt:2022zxs, Weickgenannt:2023nge, Wagner:2024fhf, Hu:2021pwh, Li:2020eon, Shi:2020htn, Banerjee:2024xnd, Bhadury:2024ckc}}, a formalism that extends conventional relativistic hydrodynamics~\cite{Romatschke:2007mq, Romatschke:2009im, Jaiswal:2016hex, Florkowski:2017olj} by incorporating spin degrees of freedom. A comprehensive review of the developments in fluid dynamic theories can be found in~\cite{Bhadury:2025dzh}.

It has recently been demonstrated in~\CITn{Abboud:2025qtg} that the formalism of perfect spin hydrodynamics for spin-$\onehalf$ particles, formulated within the classical description of spin, falls within the framework of divergence-type hydrodynamic theory. Moreover, the authors of~\CITn{Abboud:2025qtg} have shown that the considered approach is nonlinearly causal and stable. In this paper, we generalize these results to the case of quantum statistics and quantum description of spin. We consider four cases in total: Boltzmann and Fermi--Dirac statistics, analyzed separately for classical and quantum spin descriptions. For the Boltzmann case with the classical spin treatment, we reproduce the findings of~\CITn{Abboud:2025qtg}. For the remaining three cases, we demonstrate that the corresponding formulations of perfect spin hydrodynamics are likewise nonlinearly causal and stable. This implies that the framework is suitable for numerical simulations of spin dynamics in fluids such as quark-gluon plasma produced in relativistic heavy-ion collisions.

This paper is organized as follows. Section~\ref{sec:perf} examines the conditions under which perfect spin hydrodynamics satisfies the requirements of a divergence-type theory. In Sec.~\ref{sec:caus}, we specify the generating function for the relevant thermodynamic quantities in four distinct cases, i.e., Boltzmann vs.~Fermi--Dirac distribution with either classical or quantum spin treatment, and demonstrate that all the resulting formulations are nonlinearly causal and stable. We emphasize the nonperturbative character of our theory in Sec.~\ref{sec:nonpert}. The conclusions are presented in Sec.~\ref{sec:sum}.

{\it Notation and conventions} --- We use natural units, \mbox{$\hbar = c = k_{\rm B} = 1$}, and adopt the metric tensor \mbox{$g_{\mu\nu} = \textrm{diag}(+1,-1,-1,-1)$}. The scalar product of two four-vectors $a$ and $b$ is then given by \mbox{$a \cdot b = a^0 b^0 - \av \cdot \bv$}, where bold symbols denote three-vectors.  A colon denotes the contraction of two rank-two tensors, $a : b = a_{\mu\nu} b^{\mu\nu}$ (note the same order of indices in $a$ and $b$). A tilde over an antisymmetric rank-two tensor represents its dual, for example \mbox{${\tilde a}^{\alpha\beta} \equiv (\onehalf) \,\epsilon^{\alpha\beta\gamma\delta} a_{\gamma \delta}$}, where the Levi-Civita symbol $\epsilon^{\alpha\beta\gamma\delta}$ is defined by the convention \mbox{$\epsilon^{0123} =-\epsilon_{0123} = +1$}. The contraction $a_\mu Z^{\mu\nu} b_\nu$ is written compactly as $a \cdot Z \cdot b$. 

\section{Perfect spin hydrodynamics as a divergence-type theory \label{sec:perf}}


The four formulations of the perfect spin hydrodynamics for spin-1/2 particles considered in this work are based on the conservation laws
\begin{align}
    \partial_\lambda N^\lambda = 0,
    \qquad
    \partial_\lambda T^{\lambda\mu} = 0,
    \qquad
    \partial_\lambda S^{\lambda,\mu\nu} = 0, \label{consv.laws}
\end{align}
where $N^\lambda$ denotes the baryon current, $T^{\lambda\mu}$ the energy--momentum tensor and $S^{\lambda,\mu\nu}$ the spin tensor. The latter constitutes the spin part of the total angular momentum tensor $J^{\lambda,\mu\nu} = L^{\lambda,\mu\nu}+S^{\lambda,\mu\nu}$, where $L^{\lambda,\mu\nu}$ represents the orbital part. Conservation of the spin tensor implies that the energy--momentum tensor is symmetric.

Following the earlier work on divergence-type theories~\cite{Geroch:1990bw, Peralta-Ramos:2009srp, Peralta-Ramos:2010qdp, Abboud:2025qtg}, we introduce the multi-index notation $N^{\lambda A} \equiv (N^\lambda, T^{\lambda\mu}, S^{\lambda,\mu\nu})$, which allows for compact way of expressing the conservation laws in the form
\begin{align}
    \partial_\lambda N^{\lambda A} = 0. \label{consv-dtt}
\end{align}
We also introduce a set of Lagrange multipliers 
\begin{align}
    \zeta_A = (\zeta, \zeta_\mu, 
\zeta_{\mu\nu}) = (\xi, -\beta_\mu, \omega_{\mu\nu}), \label{LM}
\end{align}
where $\xi = \mu/T$ is the ratio of the baryon chemical potential $\mu$ to the temperature $T$, \mbox{$\beta_\mu = u_\mu/T$} depends on the hydrodynamic flow vector $u_\mu$, and $\omega_{\mu\nu} = \Omega_{\mu\nu}/T$ is the dimensionless spin polarization tensor associated with the spin chemical potential $\Omega_{\mu\nu}$. Since the hydrodynamic flow is normalized to unity, $u^2=1$, and $\omega_{\mu\nu}$ is an antisymmetric tensor, there are eleven independent parameters in total. 

In addition to the conserved currents, we specify the entropy current $S^\mu$ that satisfies the generalized thermodynamic relations
\begin{align}
    S^\lambda = \mathcal{N}^\lambda - \xi\, N^\lambda + \beta_\mu T^{\lambda\mu} - \frac{1}{2} \omega_{\mu\nu} S^{\lambda,\mu\nu}, \label{eq:entropy}
\end{align}
\begin{align}
    \dd S^\lambda = - \xi \dd N^\lambda + \beta_\mu \dd T^{\lambda\mu} - \frac{1}{2} \omega_{\mu\nu}\, \dd S^{\lambda,\mu\nu}, \label{eq:dentropy}
\end{align}
and
\begin{align}
\dd \mathcal{N}^\lambda  =  N^\lambda \dd\xi -  T^{\lambda\mu} \dd \beta_\mu   + \frac{1}{2}  S^{\lambda,\mu\nu}  \dd\omega_{\mu\nu}. \label{eq:dNcal}
\end{align}
For Boltzmann statistics, the current $\mathcal{N}^\lambda$ represents the sum of the particle and antiparticle currents and in the spinless case reduces to $\mathcal{N}^\lambda = P \beta^\lambda$, where $P$ is the equilibrium pressure.  Below, we provide the kinetic definitions of $\mathcal{N}^\lambda$ for all cases considered, noting that, unlike $N^\lambda$, it is not a conserved quantity. 

Equation~\EQn{eq:dentropy} can be interpreted as the first law of thermodynamics. Using our compact notation, we rewrite it  as
\begin{align}
    \dd S^\lambda = -\zeta_A\, \dd N^{\lambda A}, \label{eq:dentropyC}
\end{align}
Since the tensors $\omega_{\mu\nu}$ and $S^{\lambda,\mu\nu}$ are antisymmetric in the indices $\mu\nu$, the notation used on the right-hand side of \EQn{eq:dentropyC} implicitly includes a factor of $1/2$ in the summation over these indices, thereby avoiding the double counting. Equation~\EQn{eq:dNcal} is the generalization of the Gibbs--Duhem relation of standard thermodynamics that follows directly from~\EQn{eq:entropy} and~\EQn{eq:dentropy}. In compact form, it reads
\begin{align}
    \dd \mathcal{N}^\lambda = N^{\lambda A}\, \dd \zeta_A. \label{dchi^l}
\end{align}

In all four cases considered, we can introduce a scalar generating function $\chi$, such that the current $\mathcal{N}^\lambda$ is obtained as its derivative with respect to $\zeta_\lambda$,
\begin{align}
    \mathcal{N}^\lambda =  \frac{\partial \chi}{\partial \zeta_{\lambda}}=-\frac{\partial\chi}{\partial\beta_\lambda}.\label{eq:calNgendef}
\end{align}
From Eqs.~\eqref{dchi^l} and \eqref{eq:calNgendef}, we can express the conserved currents $N^{\lambda A}$ in terms of $\mathcal{N}^\lambda$ and~$\chi$,
\begin{align}
    N^{\lambda A} = \left(\frac{\partial \mathcal{N}^\lambda}{\partial \zeta_A}\right)
    = {-} \left(\frac{\partial^2\chi}{\partial \zeta_A \partial \beta_\lambda}\right). \label{N^lA-def}
\end{align}
Treating the fields $\zeta_A$ as independent variables, we can write $\dd N^{\lambda A} = M^{\lambda AB} d \zeta_B$, and therefore rewrite the conservation laws \eqref{consv-dtt} in the form
\begin{align}
    M^{\lambda AB} \partial_\lambda \zeta_B  = 0, \label{consv.2}
\end{align}
where
\begin{align}
    M^{\lambda AB} = \left(\frac{\partial N^{\lambda A}}{\partial\zeta_B}\right)
    = - \left(\frac{\partial^3 \chi}{\partial\zeta_B \partial\zeta_A \partial\beta_\lambda}\right). \label{M^lAB-def1}
\end{align}
It follows from~\EQn{M^lAB-def1} that $M^{\lambda AB}$ is symmetric in the indices $A$ and $B$. This object plays a central role in assessing the causality and stability of the hydrodynamic theory based on the conservation laws. These conditions are satisfied if the four-vector
\begin{align}
    M^\lambda \equiv M^{\lambda AB} Z_A Z_B
    \label{eq:Mlambda}
\end{align}
defined for any set of nonvanishing real elements $Z_A = (Z, Z_\mu, Z_{\mu\nu})$ is future-oriented and \mbox{timelike}. In this case, the hydrodynamic equations \eqref{consv.2} can be written in symmetric hyperbolic form, implying that they are nonlinearly causal and stable~\cite{Geroch:1990bw, Gavassino:2022roi, Abboud:2025qtg,GEROCH1991394}. 

To proceed further, we express the four-vector $M^\lambda$ as
\begin{align}
    M^\lambda = M^{\lambda AB}Z_AZ_B = \left( Z\frac{\partial}{\partial\xi} - Z_{\mu}\frac{\partial}{\partial\beta_{\mu}} + \frac{1}{2} Z_{\mu\nu} \frac{\partial}{\partial \omega_{\mu\nu}}\right)^2 {\cal N}^\lambda. \label{eq:MlambdaGEN}
\end{align}
The structure of the right-hand side of the equation above suggests the introduction of a linear differential operator
\begin{align}
 {\hat Z} =  Z\frac{\partial}{\partial\xi} - Z_{\mu}\frac{\partial}{\partial\beta_{\mu}} + \frac{1}{2} Z_{\mu\nu} \frac{\partial}{\partial \omega_{\mu\nu}}. \label{eq:Lhat}
\end{align}
Below, we consider different cases, distinguished by the definitions of the generating function.

\section{Nonlinear causality and stability criterion \label{sec:caus}}


\subsection{Classical spin description and Boltzmann statistics}

In this case, originally analyzed in~\cite{Abboud:2025qtg}, the generating function takes the form
\begin{align}\begin{split}
\label{eq:GFBcl}
\chi &= \int \dd P  \int \dd S \left[ f^{+_{\rm }}_{\rm eq}(x,p,s) +f^{-_{\rm }}_{\rm eq}(x,p,s)  \right],
\end{split}\end{align}
where $f^{\pm_{\rm }}_{\rm eq}(x,p,s) $ are the Boltzmann equilibrium distribution functions
\bel{eq:fpm-Bcl}
f^{\pm_{\rm }}_{\rm eq}(x,p,s) = 
\exp \LB
\pm \xi(x) - p \cdot \beta(x)  +  \frac{1}{2} \, \omega(x) : s \RB .
\eel 
 Here, the particle phase space has been extended to include the spin four-vector $s^\mu$, which is orthogonal to the particle four-momentum, $p \cdot s = 0$, and satisfies the normalization condition $s^2 = -\spin^2$, where $\spin^2 = \threefourths$ is the eigenvalue of the Casimir operator for the SU(2) group. Following the works by Mathisson~\CITn{Mathisson:1937zz,2010GReGr..42.1011M}, we define the internal angular momentum of a particle by the equation $ s^{\alpha \beta} = (1/m) \epsilon^{\alpha\beta\gamma\delta} p_\gamma s_\delta$. In the particle rest frame~(PRF), where $p^\mu = (m,0,0,0)$, the four-vector~$s^\alpha$ has only spatial components, $s^\alpha = (0,\sv_*)$, with the normalization~$\spinl = \spin$.

The integration measure in the spin space is given by~\CITn{Florkowski:2018fap}
\begin{eqnarray}
\int \dd S \ldots = \f{m}{\pi \spin}  \, \int \dd^4s \, \delta(s \cdot s + \spin^2) \, \delta(p \cdot s) \ldots\, .
\label{eq:measureS}
\end{eqnarray}
The two delta functions enforce the normalization and orthogonality conditions. The prefactor~$m/(\pi \spin)$ is chosen so that the integration reproduces the correct spin degeneracy for spin-$\onehalf$ particles,
\begin{eqnarray}
\int \dd S = \f{m}{\pi \spin}  \int \, \dd^4s \, \delta(s \cdot s + \spin^2) \, \delta(p \cdot s) = 2.
\label{eq:intS}
\end{eqnarray}

In the momentum space, the integration measure, denoted by $\dd P$, is defined as
\begin{eqnarray}\label{eq:measureP}
\int \dd P \ldots = \int \frac{\dd^3p}{(2 \pi )^3 E_p} \ldots,
\end{eqnarray}
where $E_p=\sqrt{\pv^2+m^2}$ is the particle dispersion relation. We note that both $\dd P$ and $\dd S$ are Lorentz invariant. Using~\EQ{eq:calNgendef}, we obtain the particle current
\begin{align}\begin{split}
\label{eq:calNBcl}
{\cal N}^\lambda &= \int \dd P  \int \dd S \, p^\lambda \left[ f^{+_{\rm }}_{\rm eq}(x,p,s) +f^{-_{\rm }}_{\rm eq}(x,p,s)  \right].
\end{split}\end{align}
By inserting~\EQn{eq:calNBcl} into the right-hand side of~\EQn{eq:MlambdaGEN} we arrive at
\begin{align}\begin{split}
\label{eq:MlambdaBcl}
M^\lambda &= \int \dd P  \int \dd S \, p^\lambda \left[ \left( 
Z - Z \cdot p + \frac{1}{2} Z : s 
\right)^2 f^{+}_{\rm eq}
+
\left( 
- Z - Z \cdot p + \frac{1}{2} Z : s 
\right)^2 f^{-}_{\rm eq}
\right].
\end{split}\end{align}
This result has been first obtained in~\CITn{Abboud:2025qtg}. We note that $M^\lambda$ is future-oriented and timelike, as it is a sum of the particle four-momenta $p^\lambda$ weighted by a positive factor (given by the expression in the square brackets).

\subsection{Classical spin description and Fermi--Dirac statistics}

To define the generating function for the Fermi--Dirac case, we first introduce the equilibrium Fermi--Dirac distribution functions
for particles ($+$) and antiparticles ($-$)~\CITn{Drogosz:2024gzv}
\bel{eq:fpm-FD}
f^{\pm_{\rm }}_{\rm eq}(x,p,s) = \LSB
\exp \LB
\mp \xi(x) + p \cdot \beta(x)  -  \frac{1}{2} \, \omega(x) : s \RB +1 \RSB^{-1}.
\eel 
For notational simplicity, we use here the same symbol for the equilibrium distribution function, even though it now has the Fermi--Dirac form, whereas in the previous section it denoted the Boltzmann distribution. In the following, we also adopt the compact notation
\bel{eq:fpm-FDy}
f^{\pm_{\rm }}_{\rm eq} = f_{\rm eq}(y^\pm) 
= \f{1}{e^{y^\pm} + 1}
\eel 
with
\bel{eq:ypm}
y^\pm(x,p,s) = \mp \xi(x) + p \cdot \beta(x) 
- \frac{1}{2} \omega(x) : s.
\eel

In this case, the generating function can be defined as
\begin{align}\begin{split}
\label{eq:GFFDcl}
\chi &= \int \dd P  \int \dd S \left[ F_{\rm eq}(y^+) +F^{-_{\rm }}_{\rm eq}(y^-)  \right],
\end{split}\end{align}
where the function $F_{\rm eq}(y)$ satisfies the condition
\bel{eq:Fofy}
\frac{\dd F_{\rm eq}(y)}{\dd y} = \ln( 1-  f_{\rm eq}(y)) .
\eel
The explicit expression for $F_{\rm eq}(y)$ is\footnote{The integration constant $\pi^2/6$ is chosen so that $\dd F_{\rm eq}(y)$ vanishes exponentially as $y \to \infty$.}
\bel{eq:Fexp}
F_{\rm eq}(y) = \frac{y^2}{2} + \hbox{Li}_2(-e^{y}) + \frac{\pi^2}{6},
\eel
where Li$_2(z)$ is the dilogarithm defined by the series
\bel{eq:L2}
\hbox{Li}_2(z) = \sum\limits_{k=1}^\infty \frac{z^k}{k^2}.
\eel
Equations~\EQn{eq:calNgendef} and \EQn{eq:Fofy} lead to the formula~\CITn{Drogosz:2024gzv}
\begin{align}\begin{split}
\label{eq:calNFDcl}
{\cal N}^\lambda &= -\int \dd P  \int \dd S \, p^\lambda \left[ \ln(1-f^{+}_{\rm eq}) + \ln (1-f^{-}_{\rm eq})  \right],
\end{split}\end{align}
which reduces to the Boltzmann expression in the limit of small occupation numbers, \mbox{$f^{\pm}_{\rm eq} \ll 1$}. We note that for Fermi--Dirac statistics, \EQ{eq:calNFDcl} does not correspond to the particle current \mbox{(i.e., the sum} of particle and antiparticle currents). Moreover, when spin is included, ${\cal N}^\lambda \neq P \beta^\lambda$.

Using the identities
\bel{eq:Fpm}
\frac{\dd}{\dd y} \ln\left[1 -  f_{\rm eq}(y)\right] =  f_{\rm eq}(y), \qquad \frac{\dd}{\dd y} f_{\rm eq}(y) = - f_{\rm eq}(y) \left[1-f_{\rm eq}(y)\right]
\eel
together with~\EQn{eq:MlambdaGEN}, we find 
\begin{align}\begin{split}
\label{eq:MlambdaFDcl}
M^\lambda &= \int \dd P  \int \dd S \, p^\lambda \left[ \left( 
Z + Z \cdot p + \frac{1}{2} Z : s 
\right)^2 f^{+}_{\rm eq} (1-f^+_{\rm eq}(y))
\right.
\\
& \hspace{3cm}\left.
+
\left( 
- Z + Z \cdot p + \frac{1}{2} Z : s 
\right)^2 f^{-}_{\rm eq} (1-f^-_{\rm eq}(y))
\right].
\end{split}\end{align}
Since $f^{\pm}_{\rm eq} \left(1-f^\pm_{\rm eq}(y)\right) > 0$, from~\EQ{eq:MlambdaFDcl}  we directly conclude that $M^\lambda$ is future-oriented and timelike also in the case of Fermi--Dirac statistics. 

\subsection{Quantum spin description and Boltzmann statistics}

Following our recent study~\CITn{Bhadury:2025boe}, we define the generating function for Boltzmann statistics with quantum description of spin as
\begin{align}
    \chi = 2\int \dd P \left[ f_0^+(x,p) +f_0^-(x,p) \right] \cosh\sqrt{-a^2}. \label{eq:GFBqt1}
\end{align}
Here, $f_0^\pm (x,p) = \exp(\pm \xi - \beta \cdot p)$ is the standard (spinless) Boltzmann equilibrium distribution function for particles and antiparticles, and the four-vector $a$ is defined by the formula~\cite{Bhadury:2025boe}
\bel{eq:a}
a^\mu = -\frac{1}{2m} \widetilde{\omega}^{\mu\nu} p_\nu.
\eel
We note that $a$ is spacelike, $a^2 < 0$, and the following identities hold:
\bel{eq:da}
\frac{\partial a^2}{\partial \omega_{\alpha 
\beta}}=-\frac{1}{m} \epsilon^{\alpha \beta \gamma \delta}a_{\gamma} p_\delta, \qquad
 Z_{\mu\nu}  \frac{\partial a^\alpha}{\partial \omega_{\mu\nu}} =
-\frac{1}{m } {\tilde Z}^{\alpha\beta} p_\beta, \qquad \frac{\partial \omega_{\mu \nu}}{\partial \omega_{\alpha \beta}} = g^\mu_\alpha g^\nu_\beta - g^\nu_\alpha g^\mu_\beta.
\eel
In the PRF, $a^\mu = a^\mu_* = (0, \av_*)$ with $\av_* = - \bv_*/2$. The unit vector $\nv = \av_*/\sqrt{\av_*^2}$ defines the direction of the mean spin polarization. 
Expressing the hyperbolic cosine in~\EQ{eq:GFBqt1} in terms of exponential functions, we can write the generating function in an alternative form
\bel{eq:GFBqt2}
    \chi = \sum_{i,j = \pm} \int dP \, g^{ij}_{\rm eq}(x,p), 
\eel
where $g^{ij}_{\rm eq}(x,p)$ is the Boltzmann distribution extended to spin space,
\bel{eq:gpmB}
g^{ij}_{\rm eq}(x,p) = \exp\left(i\, \xi - p \cdot \beta +j \sqrt{-a^2} \right), \quad i,j=\pm.
\eel
The first superscript of $g^{ij}_{\rm eq}(x,p)$ distinguishes particles ($+$) from antiparticles ($-$), while the second one denotes particles with spin up ($+$) and down ($-$), along the direction defined by $\nv$. Using \EQn{eq:calNgendef}, we find 
\begin{align}\begin{split}
{\cal N}^\lambda &=  \sum_{i,j = \pm} \int \dd P   \, p^\lambda  \, g^{ij}_{\rm eq}(x,p) ,
\end{split}\end{align}
which can be interpreted as the total particle current. In this case, $M^\lambda$ can again be obtained using~\EQ{eq:MlambdaGEN}; however, the square of the differential operator ${\hat Z}$ acting on ${\cal N}^\lambda$ should be calculated with care. In the first step, we find
\begin{align}
{\hat Z} \, {\cal N}^\lambda =
 \sum_{i,j = \pm} \int \dd P   \, p^\lambda  \, \left(i Z + Z \cdot p + j \frac{a \cdot {\tilde Z} \cdot p}{2 m \sqrt{-a^2}}  \right)  \, g^{ij}_{\rm eq}(x,p). 
\label{eq:firststep}
\end{align}
Acting with the differential operator ${\hat Z}$ on \EQn{eq:firststep} a second time produces two contributions. The first arises from the differentiation of the functions $g^{ij}_{\rm eq}(x,p)$ and gives
\begin{align}
 \sum_{i,j = \pm} \int \dd P   \, p^\lambda  \, \left(i Z + Z \cdot p + j \frac{a \cdot {\tilde Z} \cdot p}{2 m \sqrt{-a^2}}  \right)^2  \, g^{ij}_{\rm eq}(x,p). 
\label{eq:secondstep1}
\end{align}
The second contribution comes from the differentiation of the expression in the parentheses on the right-hand side of~\EQn{eq:firststep},
\begin{align} \frac{1}{4 m}
 \sum_{i,j = \pm} j  \int \dd P   \, p^\lambda  \,   g^{ij}_{\rm eq}(x,p) \,
 Z_{\mu\nu} \frac{\partial}{\partial \omega_{\mu\nu}}  \left(
 \frac{a \cdot {\tilde Z} \cdot p}{\sqrt{-a^2}}  \right) . 
\label{eq:secondstep2a}
\end{align}
With the help of~\EQ{eq:da}, we rewrite this term as
\begin{align} \frac{1}{4 m^2}
 \sum_{i,j = \pm} j \int \dd P   \, p^\lambda  \,   g^{ij}_{\rm eq}(x,p) \, \frac{1}{\sqrt{-a^2}}
  \left[-
 ({\tilde Z} \cdot p)^2 + \frac{(a \cdot {\tilde Z} \cdot p)^2}{a^2} \right] . 
\label{eq:secondstep2b}
\end{align}
Introducing the notation $\alpha^i = m \,\widetilde{Z}^{i0}_*$, we can write the square bracket above as $\alphav^2 - (\nv \cdot \alphav)^2$, proving its nonnegativity. Combining \EQn{eq:secondstep1} with \EQn{eq:secondstep2b}, we conclude that $M^\lambda$ can be expressed as an integral
\begin{align}
M^\lambda = \int dP \, p^\lambda \, w(x,p),
\label{eq:Mlint}
\end{align}
where $w(x,p) \geq 0$. Hence, $M^\lambda$ is again future-oriented and timelike, indicating that our theory is of the divergence type with hydrodynamic equations being nonlinearly causal and stable.

\subsection{Quantum spin description and Fermi--Dirac statistics}

In this case, following \cite{Kar:2025qvj}, we define the generating function by the formula
\bel{eq:GFFDqt}
    \chi = \sum_{i,j = \pm} \int \dd P \, G^{ij}_{\rm eq}(x,p), 
\eel
where $G^{ij}_{\rm eq} = F(y^{ij})$, with the function $F(y)$ defined by~\EQn{eq:Fofy}, and
\bel{eq:yij}
y^{ij}(x,p) = -i \, \xi(x) + p \cdot \beta(x) 
-j \, \sqrt{-a^2(x,p)}.
\eel
This directly leads to the expression
\begin{align}\begin{split}
{\cal N}^\lambda &= - \sum_{i,j = \pm} \int \dd P   \, p^\lambda  \ln(1-g^{ij}_{\rm eq}),
\end{split}\end{align}
where $g^{ij}_{\rm eq}$ have the form of the Fermi--Dirac distribution for particles and antiparticles ($i =\pm$) with spin up and down ($j=\pm$) along the direction of $\nv$,
\bel{eq:gpmFD}
g^{ij}_{\rm eq} = \left[ \exp\left(-i \, \xi + p \cdot \beta -j  \sqrt{-a^2} \right) + 1 \right]^{-1} = g_{\rm eq}(y^{ij}).
\eel
On the right-hand side of this equation, we emphasize that the functions $g^{ij}_{\rm eq}$ depend solely on the variables $y^{ij}$, simplifying the calculation of the derivatives.
Similarly to the Boltzmann case, we then find
\begin{align}
{\hat Z} \, {\cal N}^\lambda =
 \sum_{i,j = \pm} \int \dd P   \, p^\lambda  \, \left(i Z + Z \cdot p + j \frac{a \cdot {\tilde Z} \cdot p}{2 m \sqrt{-a^2}}  \right)  \, g^{ij}_{\rm eq}. 
\label{eq:firststep_FD}
\end{align}
The second application of the operator,  ${\hat Z}^2 \, {\cal N}^\lambda$, yields two terms
\begin{align}
 \sum_{i,j = \pm} \int \dd P   \, p^\lambda  \, \left(i Z + Z \cdot p + j \frac{a \cdot {\tilde Z} \cdot p}{2 m \sqrt{-a^2}}  \right)^2  \, g^{ij}_{\rm eq} \, (1-g^{ij}_{\rm eq}),
\label{eq:secondstep1_FD}
\end{align}
and
\begin{align} \frac{1}{4 m^2}
 \sum_{i,j = \pm} j \int \dd P   \, p^\lambda  \,  g^{ij}_{\rm eq}(x,p) \, \frac{1}{\sqrt{-a^2}}
  \left(-
 ({\tilde Z} \cdot p)^2 + \frac{(a \cdot {\tilde Z} \cdot p)^2}{a^2} \right) . 
\label{eq:secondstep2b_FD}
\end{align}
These results again lead to the expression \EQn{eq:Mlint} with nonnegative $w$. Hence, the spin hydrodynamics constructed in this case is also causal and stable.

\section{Nonperturbative character of the theory
\label{sec:nonpert}}

In all of the cases considered above, it was crucial to assume that the distribution functions, as well as certain combinations of them, are positive. This property may be violated if the distribution functions are expanded in the components of the spin polarization tensor. For example, in the case of classical spin and Boltzmann statistics, one may naturally use the formula
\bel{eq:fpm-Bcl-approx}
f^{\pm_{\rm }}_{\rm eq}(x,p,s) &=& 
\exp \LB
\pm \xi - p \cdot \beta  +  \frac{1}{2} \, \omega : s \RB \nonumber \\
&\approx&
\exp \LB
\pm \xi - p \cdot \beta  \RB
\LB
1 + \frac{1}{2} \, \omega : s +\cdots \RB.
\eel 
As a matter of fact, such expansions are commonly used, since spin-polarization effects are expected to be small and linear terms in $\omega$ to be sufficient. The expression in the round brackets on the right-hand side of \EQn{eq:fpm-Bcl-approx}, when truncated to a finite number of terms, may become negative if $\omega$ becomes too large.  

Consequently, in cases where expansions in $\omega$ are used \cite{Drogosz:2025iyr}, the assumptions of this work may no longer hold, and one may encounter instabilities. Such behavior may explain why numerical solutions diverge in simulations when the initial values of the spin polarization tensor are large~\cite{Drogosz:2024lkx}. On the other hand, our results hold if the exact expressions for the distribution functions are used, which may be interpreted as their nonperturbative character.

We note also that the theory is defined under the assumption that the integrals appearing in the generating functions (\ref{eq:GFBcl}), ({\ref{eq:GFFDcl}), (\ref{eq:GFBqt1}), (\ref{eq:GFFDqt}) converge.
For classical and quantum spin descriptions with Boltzmann statistics, this was studied in Ref.~\cite{Drogosz:2025ihp}, where the limit on components of $\omega$ for given particle mass, temperature, and
hydrodynamic flow was derived and found to be sufficiently high to cover the parameter range typical of heavy-ion collisions. The applicability range is the same in the case of Fermi--Dirac statistics, as mentioned in Ref.~\cite{Kar:2025qvj}, owing to the identical asymptotic behavior of the relevant expressions in the high-momentum regime, decisive for integral convergence.

\section{Summary
\label{sec:sum}}

In this work, we analyze the stability and causality of the perfect spin hydrodynamics for \mbox{spin-$\onehalf$} particles. Using the formalism developed in Ref.~\cite{Abboud:2025qtg}, we confirm that, for a Boltzmann distribution with classical spin description, perfect spin hydrodynamics fits within the framework of divergence-type theories and is nonlinearly causal and stable. We further generalize this result by considering classical phase-space distribution with a quantum treatment of spin, as well as the Fermi--Dirac statistics incorporating spin in either the classical or the quantum way. By~constructing and analyzing the generating functions for the relevant thermodynamic currents, we conclude that each of the formulations considered corresponds to a divergence-type hydrodynamic theory that is nonlinearly causal and stable. These properties ensure that the theory yields physically meaningful, numerically consistent predictions for any dynamical description of strongly interacting matter. We emphasize that the theory has a nonperturbative character, i.e., the criteria hold if the exact expressions for the distribution functions are used.

\medskip
\begin{acknowledgments}

{\it Acknowledgments:} This work was supported in part by the National Science Centre, Poland (NCN) Grant No.~2022/47/B/ST2/01372. We thank Lorenzo Gavassino and Rajeev Singh for useful comments.
\end{acknowledgments}

\end{document}